\documentstyle[amstex,amsfonts,12pt]{article}

\catcode`\@=11
\def\numberbysection{\@addtoreset{equation}{section}
        \def\theequation{\thesection.\arabic{equation}}}
\numberbysection

\begin{document}

\newlength{\lno} \lno0.5cm \newlength{\len} \len=\textwidth%
\addtolength{\len}{-\lno}

\setcounter{page}{0}

\baselineskip7mm \newpage \setcounter{page}{0}

\begin{titlepage}     
\vspace{1.5cm}
\begin{center}
{\Large\bf  $A_{n-1}^{(1)} $ Reflection K-Matrices  }\\
\vspace{1cm}
{\large A. Lima-Santos }\footnote{e-mail: dals@df.ufscar.br} \\
\vspace{1cm}
{\large \em Universidade Federal de S\~ao Carlos, Departamento de F\'{\i}sica \\
Caixa Postal 676, CEP 13569-905~~S\~ao Carlos, Brasil}\\
\end{center}
\vspace{2.5cm}

\begin{abstract}
We investigate the possible regular solutions of the boundary Yang-Baxter
equation for the  vertex models associated with the $A_{n-1}^{(1)}$ affine Lie algebra.
 We have classified them in two classes of solutions. The first class consists of
 $n(n-1)/2$  K-matrix solutions with three free parameters. The second class are solutions that depend on the  parity of $n$.  
For $n$ odd there exist $n$ reflection K-matrices with $2+[n/2]$ free parameters.
It turns out that for  $n$ even there exist  $n/2$ $K$-matrices with $2+n/2$ free parameters and $n/2$ $K$-matrices
with $1+n/2$ free parameters.

\end{abstract}
\vfill
\begin{center}
\small{\today}
\end{center}
\end{titlepage}

\baselineskip6mm

\newpage{}

\section{{}Introduction}

The search for integrable models through the Yang-Baxter equation \cite%
{Baxter, KIB, ABR}%
\begin{equation}
R_{12}(u-v)R_{13}(u)R_{23}(v)=R_{23}(v)R_{13}(u)R_{12}(u-v)  \label{int.1}
\end{equation}%
has been performed by the quantum group approach in \cite{KR}. Thus the
problem is reduced to a linear one. Indeed, $R$ matrices corresponding to
vector representations of all non-exceptional affine Lie algebras were
determined in this way in \cite{Jimbo}.

A similar approach is clearly desirable for finding solutions $K(u)$ of the
boundary Yang-Baxter equation \cite{Cherednik, Sklyanin}%
\begin{equation}
R_{12}(u-v)K_{1}(u)R_{21}(u+v)K_{2}(v)=K_{2}(v)R_{12}(u+v)K_{1}(u)R_{21}(u-v).
\label{int.2}
\end{equation}%
With this goal in mind, the study of boundary quantum groups was initiated
in \cite{MN}. However, as observed by Nepomechie \cite{Nepo}, an independent
systematic method of constructing the boundary quantum group generators is
not yet available. In contrast to the bulk case \cite{Jimbo}, one cannot
exploit boundary affine Toda field theory, since appropriated classical
integrable boundary conditions are not yet known \cite{BCDR}.

We are also sharing the hope that by studying the known examples of boundary
quantum group generators, it may become possible to uncover their basic
algebraic structure, and to find generalizations to all affine Lie algebras.
Independent of the lack of an algebraic solution from the quantum group
approach, there has been an increasing amount of effort towards the
understanding of two-dimensional integrable theories with reflecting
boundaries via solutions of the reflection equation (\ref{int.2}). In field
theory, attention is focused on the boundary $S$-matrix. In statistical
mechanics, the emphasis has been on deriving solutions of (\ref{int.2}) and
the calculation of various surface critical phenomena, both at and away from
criticality \cite{Batchelor}. In condensed matter physics the actual target
is the impurity problem.

The classification of all possible solutions of the reflection equation (\ref%
{int.2}) by direct computation has been seen as a very difficult problem.
However, recently we have proposed a method which allows the classification
of the $D_{n+1}^{(2)}$ reflection $K$-matrices \cite{Lima2} as well as the $%
K $-matrices of the $19$-vertex models \cite{Lima1}. \ In spite of these
papers we decided to continue in this line in order to include the $%
A_{n-1}^{(1)}$ reflection $K$-matrices which will reveal us its algebraic
structure.

We have organized this paper as follows. In Section $2$ we choose the $%
A_{n-1}^{(1)}$ reflection equations \ and in Section $3$ their solutions are
derived and classified in two types. The last section is reserved for the
conclusion. The first models have its $K$-matrices written explicitly in
appendices.

\section{The A$_{n-1}^{(1)}$ Reflection Equations}

The $R$-matrix for the vertex models associated with the $A_{n-1}^{(1)}$ ($%
n\geq 2$ ) affine Lie algebra was originally found in the articles \cite%
{Cherednik1, BVV} and as presented in \cite{Jimbo} it has the form 
\begin{eqnarray}
R(u) &=&a_{1}(u)\sum E_{ii}\otimes E_{ii}+a_{2}(u)\sum_{i\neq
j}E_{ii}\otimes E_{jj}  \nonumber \\
&&+a_{3}(u)\sum_{i<j}E_{ij}\otimes E_{ji}+a_{4}(u)\sum_{i>j}E_{ij}\otimes
E_{ji},  \label{re.1}
\end{eqnarray}%
where $E_{ij}$ denotes the elementary $n$ by $n$ matrices ($%
(E_{ij})_{ab}=\delta _{ia}\delta _{ib}$) and the Boltzmann weights with
functional dependence on the spectral parameter $u$ are given by 
\begin{equation}
a_{1}(u)=({\rm e}^{u}-q^{2}),\ a_{2}(u)=q({\rm e}^{u}-1),\
a_{3}(u)=-(q^{2}-1),\ a_{4}(u)=-{\rm e}^{u}(q^{2}-1).  \label{re.2}
\end{equation}%
Here $q$ denotes an arbitrary parameter.

For $n>2$, the $R$-matrix (\ref{re.1}) does not enjoy {\rm P} and {\rm T}
symmetry but just {\rm PT} invariance%
\begin{equation}
{\cal P}_{12}R_{12}(u){\cal P}_{12}\equiv R_{21}(u)=R_{12}(u)^{t_{1}t_{2}}
\label{re.3}
\end{equation}%
and unitarity%
\begin{equation}
R_{12}(u)R_{21}(-u)=\zeta (u)=a_{1}(u)a_{1}(-u).  \label{re.4}
\end{equation}%
It is not crossing invariant either but it obeys the weaker property \cite%
{RSTS}%
\begin{equation}
\left\{ \left\{ \left\{ R_{12}(u)^{t_{2}}\right\} ^{-1}\right\}
^{t_{2}}\right\} ^{-1}=\frac{\zeta (u+\rho )}{\zeta (u+2\rho )}%
M_{2}R_{12}(u+2\rho )M_{2}^{-1},  \label{re.5}
\end{equation}%
where $M$ is a symmetry of the $R$-matrix 
\begin{equation}
\left[ R(u),M\otimes M\right] =0,\quad M_{ij}=\delta _{ij}q^{n+1-2i},\quad
\rho =n\ln q.  \label{re.6}
\end{equation}

The matrix $K_{-}(u)$ satisfies the left boundary Yang-Baxter equation, also
known as the reflection equation,%
\begin{equation}
R_{12}(u-v)\overset{1}{K_{-}}(u)R_{21}(u+v)\overset{2}{K_{-}}(v)=\overset{2}{%
K_{-}}(v)R_{12}(u+v)\overset{1}{K_{-}}(u)R_{21}(u-v),  \label{re.7}
\end{equation}%
which governs the integrability at boundary for a given bulk theory. A
similar equation should also hold for the matrix $K_{+}(u)$ at the opposite
boundary. However, for the $A_{n-1}^{(1)}$ models, one can see from \cite%
{Nepo2} that the corresponding quantity 
\begin{equation}
K_{+}(u)=K_{-}(-u-\rho )^{{\rm t}}M,  \label{re.8}
\end{equation}%
satisfies the right boundary Yang-Baxter equation. Here ${\rm t}={\rm t}_{1}%
{\rm t}_{2}$ and {\rm t}$_{i}$ stands for transposition taken in the $i^{th}$
space.

Therefore, we can start for searching the matrices $K_{-}(u)$. In this paper
only regular solutions will be considered, although there is much interest
for non-regular $A_{n-1}^{(1)}$ solutions \cite{Ganden, Delius}.

Regular solutions mean that the matrix $K_{-}(u)$ has the form 
\begin{equation}
K_{-}(u)=\sum_{i,j=1}^{n}k_{ij}(u)\ E_{ij}  \label{re.9}
\end{equation}%
and satisfies the condition 
\begin{equation}
k_{ij}(0)=\delta _{ij},\quad \qquad i,j=1,2,...,n.  \label{re.10}
\end{equation}

Substituting (\ref{re.1}) and (\ref{re.9}) into (\ref{re.7}), we will get $%
n^{4}$ functional equations for the $k_{ij}$ matrix elements, many of which
are dependent. In order to solve them, we shall proceed in the following
way. First we consider the $(i,j)$ component of the matrix equation (\ref%
{re.7}). By differentiating it with respect to $v$ and taking $v=0$, we get
algebraic equations involving the single variable $u$ and $n^{2}$ parameters 
\begin{equation}
\beta _{ij}=\frac{dk_{ij}(v)}{dv}|_{v=0}\qquad i,j=1,2,...,n.  \label{re.11}
\end{equation}%
Second, these algebraic equations are denoted by $E[i,j]=0$ and collected
into blocks $B[i,j]$ , $i=1,...,n^{2}-i$ and $j=i,i+1,...,n^{2}-i$, defined
by 
\begin{equation}
B[i,j]=\left\{ 
\begin{array}{c}
E[i,j]=0,\ E[j,i]=0,\  \\ 
E[n^{2}+1-i,n^{2}+1-j]=0,\ E[n^{2}+1-j,n^{2}+1-i]=0.%
\end{array}%
\right. \   \label{re.12}
\end{equation}%
For a given block $B[i,j]$, the equation $E[n^{2}+1-i,n^{2}+1-j]=0$ can be
obtained from the equation $E[i,j]=0$ by interchanging 
\begin{equation}
k_{ij}\longleftrightarrow k_{n+1-i\ n+1-j},\quad \beta
_{ij}\longleftrightarrow \beta _{n+1-i\ n+1-j^{\prime }},\quad
a_{3}(u)\leftrightarrow a_{4}(u)\quad \   \label{re.13}
\end{equation}%
and the equation $E[j,i]=0$ is obtained from the equation $E[i,j]=0$ by the
interchanging 
\begin{equation}
k_{ij}\longleftrightarrow k_{ji},\quad \beta _{ij}\longleftrightarrow \beta
_{ji}\quad  \label{re.14}
\end{equation}%
In this way, we can control all equations and a particular solution is
simultaneously connected with at least four equations.

\section{The $A_{n-1}^{(1)}$ K-Matrix Solutions}

Analyzing the $A_{n-1}^{(1)}$ reflection equations\ one can see that they
possess a very special structure. Several equations exist involving only the
elements out of the diagonal, $k_{ij}\ (i\neq j)$, these are the simplest
equations and we will solve them first.

By direct inspection one can see that the diagonal blocks $B[i,i]$ are
uniquely solved by the relations%
\begin{equation}
\beta _{ij}k_{ji}(u)=\beta _{ji}k_{ij}(u),\qquad \forall \ i\neq j
\label{sol.1}
\end{equation}%
It means that we only need to find the $n(n-1)/2$ elements $k_{ij}\ (i<j)$.
Now we choose a particular $k_{ij}$ $(i<j)$ to be different from zero, with $%
\beta _{ij}\neq 0$, and try to express all remaining elements in terms of
this particular element. We have verified that this is possible provided
that 
\begin{equation}
k_{pq}(u)=\left\{ 
\begin{array}{c}
\frac{a_{4}(u)}{a_{3}(u)}\frac{\beta _{pq}}{\beta _{ij}}k_{ij}(u)\quad {\rm %
if}\quad p>i\quad {\rm and}\quad q>j \\ 
\\ 
\frac{\beta _{pq}}{\beta _{ij}}k_{ij}(u)\quad {\rm if}\quad p>i\quad {\rm and%
}\quad q<j%
\end{array}%
\right. ,\qquad (p\neq q)  \label{sol.2}
\end{equation}%
Combining (\ref{sol.1}) with (\ref{sol.2}) we will obtain a very strong
entail for the elements out of the diagonal 
\begin{equation}
k_{ij}(u)\neq 0\Rightarrow \left\{ 
\begin{array}{c}
k_{pj}(u)=0\quad {\rm for}\quad p\neq i \\ 
\\ 
k_{iq}(u)=0\quad {\rm for}\quad q\neq j%
\end{array}%
\right.  \label{sol.3}
\end{equation}%
It means that for a given $k_{ij}$, the only elements different from zero \
in the $i^{th}$-row and in the $j^{th}$-column of $K_{-}(u)$ are $%
k_{ii},k_{ij},k_{jj}$ and $k_{ji}$.

Analyzing more carefully these equations with the conditions (\ref{sol.1})
and (\ref{sol.3}), we have found from the $n(n-1)/2$ matrix elements $%
k_{ij}\ (i<j)$ that there are two possibilities to choose a particular $%
k_{ij}\neq 0:$

\begin{itemize}
\item Only one non-diagonal element and its symmetric are allowed to be
different from zero. Thus we have $n(n-1)/2$ reflection $K$-matrices with $%
n+2$ non-zero elements. Here we will denote by ${\Bbb K}_{ij}^{I}$ ($i<j$ ),
the $K$-matrix for which the non-diagonal element $k_{ij}$ is the one chosen
to be the non-zero matrix element. These matrices will be named {\rm Type-I }%
solutions.

\item For each $k_{ij}\neq 0,$ additional non-diagonal elements and its
asymmetric are allowed to be different from zero provided they satisfy the
equations 
\begin{equation}
k_{ij}(u)k_{ji}(u)=k_{rs}(u)k_{sr}(u)\qquad {\rm with}\quad i+j=r+s\quad 
\text{{\rm mod} }n  \label{sol.4a}
\end{equation}%
It means that we will get $n$ reflection $K$-matrices with the number of
non-zero elements depending on the parity of $n.$ Next, we choose the \ $n$
possible particular elements as being $k_{1j}$ , $j=1,2,\cdots ,n$ and $%
k_{2n}.$ We will also denote the corresponding $K$-matrices by ${\Bbb K}%
_{1j}^{II}$, $j=1,2,\cdots ,n$ and ${\Bbb K}_{2n}^{II}$, respectively. These
matrices are named {\rm Type-II}{\em \ }solutions.
\end{itemize}

\bigskip For example, the $A_{2}^{(1)}$ model has only {\rm Type-I}
solutions. The $K$-matrices are%
\begin{eqnarray}
{\Bbb K}_{12}^{I} &=&\left( 
\begin{array}{ccc}
k_{11} & {\bf k}_{12} & 0 \\ 
k_{21} & k_{22} & 0 \\ 
0 & 0 & k_{33}%
\end{array}%
\right) ,{\Bbb K}_{13}^{I}=\left( 
\begin{array}{ccc}
k_{11} & 0 & {\bf k}_{13} \\ 
0 & k_{22} & 0 \\ 
k_{31} & 0 & k_{33}%
\end{array}%
\right) ,{\Bbb K}_{23}^{I}=\left( 
\begin{array}{ccc}
k_{11} & 0 & 0 \\ 
0 & k_{22} & {\bf k}_{23} \\ 
0 & k_{32} & k_{33}%
\end{array}%
\right)  \nonumber \\
&&  \label{sol.5}
\end{eqnarray}%
One can expect that these are the three possibilities to write the same
solution for the $A_{2}^{(1)}$ model.

For the $A_{3}^{(1)}$ model we have six {\rm Type-I} \ solutions $\{{\Bbb K}%
_{12}^{I},{\Bbb K}_{13}^{I},{\Bbb K}_{14}^{I},{\Bbb K}_{23}^{I},{\Bbb K}%
_{24}^{I},{\Bbb K}_{34}^{I}\}$\ all with six non-zero elements. In this
model we also have two {\rm Type-II} solutions $\{{\Bbb K}_{12}^{II},{\Bbb K}%
_{14}^{II}\}:$ 
\begin{equation}
\begin{array}{ccc}
{\Bbb K}_{12}^{II}=\left( 
\begin{array}{cccc}
k_{11} & {\bf k}_{12} & 0 & 0 \\ 
k_{21} & k_{22} & 0 & 0 \\ 
0 & 0 & k_{33} & k_{34} \\ 
0 & 0 & k_{43} & k_{44}%
\end{array}%
\right) & {\rm and} & {\Bbb K}_{14}^{II}=\left( 
\begin{array}{cccc}
k_{11} & 0 & 0 & {\bf k}_{14} \\ 
0 & k_{22} & k_{23} & 0 \\ 
0 & k_{32} & k_{33} & 0 \\ 
k_{41} & 0 & 0 & k_{44}%
\end{array}%
\right) \\ 
\begin{array}{c}
\\ 
k_{12}k_{21}=k_{34}k_{43}%
\end{array}
&  & 
\begin{array}{c}
\\ 
k_{14}k_{41}=k_{23}k_{32}%
\end{array}%
\end{array}
\label{sol.5a}
\end{equation}%
For $n\geq 5$, in addition to the $n(n-1)/2$ \ {\rm Type-I} solutions with $%
n+2$ non-zero matrix elements, we have also $n$ {\rm Type-II} solutions with
the following property: if $n$ is odd these $K$-matrices have $2n-1$
non-zero elements, but if $n$ is even, half of these $K$-matrices have $2n$
non-zero elements and the remaining ones are matrices with $2n-2$ non-zero
elements.

Although we already know as counting the $K$-matrices for the $A_{n-1}^{(1)}$
models we still have to identify among them which are similar. Indeed we can
see a ${\Bbb Z}_{n}$ similarity transformation which maps their matrix
elements positions:%
\begin{equation}
K^{(\alpha )}=h_{\alpha }K^{(0)}h_{n-\alpha },\quad \alpha =0,1,2,\cdots ,n-1
\label{sol.6}
\end{equation}%
where $h_{\alpha }$ are the ${\Bbb Z}_{n}$ matrices%
\begin{equation}
\left( h_{\alpha }\right) _{ij}=\delta _{i,i+\alpha }\qquad {\rm mod\ }n
\label{sol.7}
\end{equation}%
In order to do this we can choose $K^{(0)}$ as ${\Bbb K}_{12}^{II}$ and the
similarity transformations (\ref{sol.6}) give us the $K^{(\alpha )}$
matrices whose matrix elements are in the same positions of the matrix
elements of the ${\Bbb K}_{1j}^{II}$ and ${\Bbb K}_{2n}^{II}$ matrices.
However, due to the fact that the relations (\ref{sol.2}) involve the ratio $%
\frac{a_{4}(u)}{a_{3}(u)}={\rm e}^{u}$, as well as the additional
constraints (\ref{sol.4a}), we could not find a similarity transformation
among these ${\Bbb K}^{^{\prime }}s$ matrices, even after a gauge
transformation. Even for the {\rm Type-I} solutions the similarity account
is not simple due to the presence of three types of scalar functions and the
constraint equations for the parameters $\beta _{ij}$. Nevertheless, as we
have found a way to write all solutions, we can leave the similarity account
to the reader.

Having identified these possibilities we may proceed in order to find the $n$
diagonal elements $k_{ii}(u)$ in terms of the non-diagonal elements $%
k_{ij}(u)$ for each ${\Bbb K}_{ij}$ matrix.

These procedure is now standard \cite{Lima1}. For instance, if we are
looking for ${\Bbb K}_{12}^{II}$, the non-diagonal elements $k_{ij},$ ($%
i+j=3\quad ${\rm mod} $n$ ) \ in terms of $k_{12}$ are given by%
\begin{equation}
k_{ij}(u)=\left\{ 
\begin{array}{c}
\frac{\beta _{ij}}{\beta _{12}}k_{12}(u)\qquad \quad {\rm for}\quad \qquad
\quad i+j=3 \\ 
\\ 
\frac{\beta _{ij}}{\beta _{12}}{\rm e}^{u}k_{12}(u)\quad {\rm for}\quad
i+j=3\quad {\rm mod\ }n \\ 
\\ 
0\qquad \qquad \qquad \quad \qquad {\rm otherwise}%
\end{array}%
\right.  \label{sol.8a}
\end{equation}%
for $i,j=1,2,\cdots n,\quad (i\neq j).$

Substituting (\ref{sol.8a}) into the reflection equations we can now easily
find the $k_{ii}$ elements up to an arbitrary function, here identified as $%
k_{12}(u)$. Moreover, their consistency relations will yield us some
constraints equations for the parameters $\beta _{ij}$.

After we have found all diagonal elements in terms of $k_{ij}(u)$, we can,
without loss of generality, choose the arbitrary functions as%
\begin{equation}
k_{ij}(u)=\frac{1}{2}\beta _{ij}({\rm e}^{2u}-1),\qquad i<j.  \label{sol.9}
\end{equation}%
This choice allows us to work out the solutions in terms of the functions $%
f_{ii}(u)$ and $h_{ij}(u)$ defined by%
\begin{equation}
f_{ii}(u)=\beta _{ii}({\rm e}^{u}-1)+1\qquad {\rm and}\qquad h_{ij}(u)=\frac{%
1}{2}\beta _{ij}({\rm e}^{2u}-1),  \label{sol.10}
\end{equation}%
for $i,j=1,2,\cdots ,n.$

Now, we will simply present the general solutions and write them explicitly
for the first values of $n$ in appendices. Let us start considering the {\rm %
Type-I} solutions.

\subsection{The Type-I K-Matrices}

\bigskip Here we have $n(n-1)/2$ reflection $K$-matrices with $n+2$ non-zero
elements. For $1<i<j\leq n$ we get $(n-2)(n-1)/2$ solutions 
\begin{eqnarray}
{\Bbb K}_{ij}^{I} &=&f_{ii}(u)E_{ii}+{\rm e}%
^{2u}f_{ii}(-u)E_{jj}+h_{ij}(u)E_{ij}+h_{ji}(u)E_{ji}  \nonumber \\
&&+{\cal Z}_{i}(u)\sum_{l=1}^{i-1}E_{ll}+{\cal Y}_{i+1}^{(i)}(u)%
\sum_{l=i+1}^{j-1}E_{ll}+{\rm e}^{2u}{\cal Z}_{i}(u)\sum_{l=j+1}^{n}E_{ll},
\label{first.1}
\end{eqnarray}%
where ${\cal Z}_{i}(u)$ and ${\cal Y}_{i+1}^{(i)}(u)$ are scalar functions
defined by%
\begin{equation}
{\cal Z}_{i}(u)=f_{ii}(-u)+\frac{1}{2}\left( \beta _{ii}+\beta _{11}\right) 
{\rm e}^{-u}\left( {\rm e}^{2u}-1\right)  \label{first.2}
\end{equation}%
and%
\begin{equation}
{\cal Y}_{l}^{(i)}(u)=f_{ii}(u)+\frac{1}{2}\left( \beta _{ll}-\beta
_{ii}\right) \left( {\rm e}^{2u}-1\right) .  \label{first.3}
\end{equation}%
For $i=1$ and $1<j\leq n$ we get the $n-1$ remaining solutions%
\begin{eqnarray}
{\Bbb K}_{1j}^{I} &=&f_{11}(u)E_{11}+{\rm e}%
^{2u}f_{11}(-u)E_{jj}+h_{1j}(u)E_{1j}+h_{j1}(u)E_{j1}  \nonumber \\
&&+{\cal Y}_{2}^{(1)}(u)\sum_{l=2}^{j-1}E_{ll}+{\cal X}_{j+1}(u)%
\sum_{l=j+1}^{n}E_{ll},  \label{first.4}
\end{eqnarray}%
where a new scalar function appears,%
\begin{equation}
{\cal X}_{j+1}(u)={\rm e}^{2u}f_{11}(-u)+\frac{1}{2}\left( \beta _{j+1\
j+1}+\beta _{11}-2\right) {\rm e}^{u}\left( {\rm e}^{2u}-1\right) .
\label{first.5}
\end{equation}%
The number of free parameters is fixed by the constraint equations which
depend on the presence of these scalar functions: when ${\cal Y}%
_{l}^{(i)}(u) $ is present in ${\Bbb K}_{ij}$ we have constraint equations
of the type 
\begin{equation}
\beta _{ij}\beta _{ji}=\left( \beta _{ll}+\beta _{ii}-2\right) \left( \beta
_{ll}-\beta _{ii}\right) ,  \label{first.6}
\end{equation}%
but, when ${\cal Z}_{i}(u)$ is present the corresponding constraints are of
the type%
\begin{equation}
\beta _{ij}\beta _{ji}=\left( \beta _{11}+\beta _{ii}\right) \left( \beta
_{11}-\beta _{ii}\right) .  \label{first.7}
\end{equation}%
The presence of at least one ${\cal X}_{j+1}(u)$ yields a third type of
constraints, 
\begin{equation}
\beta _{ij}\beta _{ji}=\left( \beta _{j+1j+1}+\beta _{11}-2\right) \left(
\beta _{j+1j+1}-\beta _{11}-2\right) \text{.}  \label{first.8}
\end{equation}%
From (\ref{first.1}) and (\ref{first.4}) we can see that in each ${\Bbb K}%
_{ij}^{I}$ we have at most two scalar functions. It means that all these $%
{\Bbb K}_{ij}^{I}$ matrices are $3$-parameter solutions of the reflection
equation. \ 

Finally, we observe that the solution with $i=1$ and $j=n$ , {\it i.e}.%
\begin{equation}
{\Bbb K}_{1n}^{I}=f_{11}(u)E_{11}+{\rm e}%
^{2u}f_{11}(-u)E_{nn}+h_{1n}(u)E_{1n}+h_{n1}(u)E_{n1}+{\cal Y}%
_{2}^{(1)}(u)\sum_{l=2}^{j-1}E_{ll}  \label{first.9}
\end{equation}%
has the constraint 
\begin{equation}
\beta _{1n}\beta _{n1}=\left( \beta _{22}+\beta _{11}-2\right) \left( \beta
_{22}-\beta _{11}\right)  \label{first.10}
\end{equation}%
and, it is the solution derived by Abad and Rios \cite{Abad}.

\subsection{The Type-II K-Matrices}

Due to the property (\ref{sol.4a}) we have found three {\rm Type-II} general
solutions for each $A_{n-1}^{(1)}$ model:%
\begin{eqnarray}
{\rm Type-IIa} &=&\left\{ {\Bbb K}_{12p}^{II}\right\} ,\quad {\rm Type-IIb}%
=\left\{ {\Bbb K}_{12p+1}^{II}\right\} ,\quad {\rm Type-IIc}={\Bbb K}%
_{2n}^{II}  \nonumber \\
p &=&1,2,\cdots ,[\frac{n}{2}]  \label{second.1}
\end{eqnarray}%
where $[\frac{n}{2}]$ being the integer part of $\frac{n}{2}.$

For $n$-odd, the {\rm Type-IIa} solution is%
\begin{eqnarray}
{\Bbb K}_{12p}^{II} &=&f_{11}(u)\sum_{j=1}^{p}E_{jj}+{\rm e}%
^{2u}f_{11}(-u)\sum_{j=p+1}^{[\frac{n}{2}]+p}E_{jj}+{\rm e}%
^{2u}f_{11}(u)\sum_{j=[\frac{n}{2}]+p+2}^{n}E_{jj}  \nonumber \\
&&+{\cal X}_{[\frac{n}{2}]+p+1}(u)E_{[\frac{n}{2}]+p+1\ [\frac{n}{2}%
]+p+1}+\left( \sum\begin{Sb} i+j=1+2p\  \\ i\neq j  \end{Sb}  +\sum\begin{Sb}
i+j=1+2p\ {\rm mod\ }n  \\ i\neq j  \end{Sb}  {\rm e}^{u}\right)
h_{ij}(u)E_{ij},  \nonumber \\
&&  \label{second.2}
\end{eqnarray}%
with constraint equations%
\begin{eqnarray}
\beta _{rs}\beta _{sr} &=&\left( \beta _{\lbrack \frac{n}{2}]+p+1\ [\frac{n}{%
2}]+p+1}+\beta _{11}-2\right) \left( \beta _{\lbrack \frac{n}{2}]+p+1\ [%
\frac{n}{2}]+p+1}-\beta _{11}-2\right)  \nonumber \\
r+s &=&1+2p\quad \text{{\rm mod} }n.  \label{second.3}
\end{eqnarray}%
For the {\rm Type-IIb} solutions we have obtained the following matrices

\begin{eqnarray}
{\Bbb K}_{12p+1}^{II} &=&f_{11}(u)\sum_{j=1}^{p}E_{jj}+{\rm e}%
^{2u}f_{11}(-u)\sum_{j=p+2}^{[\frac{n}{2}]+p+1}E_{jj}+{\rm e}%
^{2u}f_{11}(u)\sum_{j=[\frac{n}{2}]+p+2}^{n}E_{jj}  \nonumber \\
&&+{\cal X}_{p+1}(u)E_{p+1\ p+1}+\left( \sum\begin{Sb} i+j=2+2p\  \\ i\neq j 
\end{Sb}  +\sum\begin{Sb} i+j=2+2p\ {\rm mod\ }n  \\ i\neq j  \end{Sb}  {\rm %
e}^{u}\right) h_{ij}(u)E_{ij},  \nonumber \\
&&  \label{second.4}
\end{eqnarray}%
together with their constraint equations 
\begin{eqnarray}
\beta _{rs}\beta _{sr} &=&\left( \beta _{p+1\ p+1}+\beta _{11}-2\right)
\left( \beta _{p+1\ p+1}-\beta _{11}\right)  \nonumber \\
r+s &=&2+2p\quad \text{{\rm mod} }n.  \label{second.5}
\end{eqnarray}%
Finally, the {\rm Type-IIc} solution is the matrix ${\Bbb K}_{2n}^{II}$ 
\begin{eqnarray}
{\Bbb K}_{2n}^{II} &=&{\cal Z}_{2}(u)E_{11}+{\rm e}^{2u}f_{22}(-u)%
\sum_{j=2}^{[\frac{n}{2}]+1}E_{jj}+{\rm e}^{2u}f_{22}(u)\sum_{j=[\frac{n}{2}%
]+2}^{n}E_{jj}  \nonumber \\
&&+\sum\begin{Sb} i+j=2\ \text{{\rm mod} }n  \\ i\neq j  \end{Sb} 
h_{ij}(u)E_{ij},  \label{second.6}
\end{eqnarray}%
for which the constraint equations are 
\begin{eqnarray}
\beta _{rs}\beta _{sr} &=&\left( \beta _{11}+\beta _{22}\right) \left( \beta
_{11}-\beta _{22}\right)  \nonumber \\
r+s &=&2\quad \text{{\rm mod} }n.  \label{second.7}
\end{eqnarray}%
The function $\ {\cal Z}_{2}(u)$ is given by (\ref{first.2}) and the
functions ${\cal X}_{j}(u)$ by (\ref{first.5}), while the functions $%
f_{11}(u)$ , $f_{22}(u)$ and $h_{ij}(u)$ are given by (\ref{sol.10}).
Therefore we have $n$ reflection $K$-matrices for the $A_{n-1}^{(1)}$ models
($n$ odd). They are $(2+[\frac{n}{2}])$ -free parameter solutions with $2n-1$
non-zero matrix elements.

When $n$ is even we have a similar identification but substantial
differences exist.

In the $n$-even case the {\rm Type-IIa} solutions are the matrices%
\begin{eqnarray}
{\Bbb K}_{12p}^{II} &=&f_{11}(u)\sum_{j=1}^{p}E_{jj}+{\rm e}%
^{2u}f_{11}(-u)\sum_{j=p+1}^{\frac{n}{2}+p}E_{jj}+{\rm e}^{2u}f_{11}(u)%
\sum_{j=\frac{n}{2}+p+1}^{n}E_{jj}  \nonumber \\
&&+\left( \sum\begin{Sb} i+j=2\  \\ i\neq j  \end{Sb}  +\sum\begin{Sb} %
i+j=1+2p\ \text{{\rm mod} }n  \\ i\neq j  \end{Sb}  {\rm e}^{u}\right)
h_{ij}(u)E_{ij},  \label{second.10}
\end{eqnarray}%
with constraint equations%
\begin{equation}
\beta _{12p}\beta _{2p1}=\beta _{rs}\beta _{sr},\qquad r+s=1+2p\quad {\rm %
mod\ }n.  \label{second.11}
\end{equation}%
For the {\rm Type-IIb} solutions we have%
\begin{eqnarray}
{\Bbb K}_{12p+1}^{II} &=&f_{11}(u)\sum_{j=1}^{p}E_{jj}+{\cal Y}%
_{p+1}^{(1)}(u)E_{p+1\ p+1}+{\rm e}^{2u}f_{11}(-u)\sum_{j=p+2}^{\frac{n}{2}%
+p}E_{jj}  \nonumber \\
&&+{\cal X}_{\frac{n}{2}+p+1}(u)E_{\frac{n}{2}+p+1\ \frac{n}{2}+p+1}+{\rm e}%
^{2u}f_{11}(u)\sum_{j=\frac{n}{2}+p+2}^{n}E_{jj}  \nonumber \\
&&+\left( \sum\begin{Sb} i+j=2+2p\  \\ i\neq j  \end{Sb}  +\sum\begin{Sb} %
i+j=2+2p\ {\rm mod\ }n  \\ i\neq j  \end{Sb}  {\rm e}^{u}\right)
h_{ij}(u)E_{ij},  \label{second.12}
\end{eqnarray}%
with the following constraint equations%
\begin{eqnarray}
\beta _{rs}\beta _{sr} &=&\left( \beta _{p+1\ p+1}+\beta _{11}-2\right)
\left( \beta _{p+1\ p+1}-\beta _{11}\right)  \nonumber \\
&=&\left( \beta _{\frac{n}{2}+p+1\ \frac{n}{2}+p+1}+\beta _{11}-2\right)
\left( \beta _{\frac{n}{2}+p+1\ \frac{n}{2}+p+1}-\beta _{11}-2\right) , 
\nonumber \\
r+s &=&2+2p\quad {\rm mod\ }n,  \label{second.13}
\end{eqnarray}%
Again, the {\rm Type-IIc }solution is the matrix ${\Bbb K}_{2n}^{II}$ 
\begin{eqnarray}
{\Bbb K}_{2n}^{II} &=&{\cal Z}_{2}(u)E_{11}+f_{22}(u)\sum_{j=2}^{\frac{n}{2}%
}E_{jj}+{\cal Y}_{\frac{n}{2}+1}^{(2)}(u)E_{\frac{n}{2}+1\ \frac{n}{2}+1} 
\nonumber \\
&&+{\rm e}^{2u}f_{22}(-u)\sum_{j=\frac{n}{2}+2}^{n}E_{jj}+\sum\begin{Sb} %
i+j=2\ {\rm mod\ }n  \\ i\neq j  \end{Sb}  h_{ij}(u)E_{ij},
\label{second.14}
\end{eqnarray}%
with the constraint equations%
\begin{eqnarray}
\beta _{rs}\beta _{sr} &=&\left( \beta _{11}-\beta _{22}\right) \left( \beta
_{11}-\beta _{22}\right)  \nonumber \\
&=&\left( \beta _{_{\frac{n}{2}+1\ \frac{n}{2}+1\ }}+\beta _{22}-2\right)
\left( \beta _{_{\frac{n}{2}+1\ \frac{n}{2}+1\ }}-\beta _{22}\right) , 
\nonumber \\
r+s &=&2\quad {\rm mod\ }n.  \label{second.15}
\end{eqnarray}%
where the scalar functions ${\cal Z}_{2}(u)$ and ${\cal Y}_{p+1}^{(2)}(u)$
are given by (\ref{first.2}) and (\ref{first.3}), respectively.

Here we observe that for $n$ even, the {\rm Type-IIa} is a $(2+\frac{n}{2})$%
-free parameter solution with $2n$ non-zero matrix elements, while the {\rm %
Type-IIb} and the {\rm Type-IIc} are $(1+\frac{n}{2})$-free parameter
solutions with $2(n-1)$ non-zero matrix elements.

\section{Conclusion}

The absence of an algebraic method such as the quantum group approaches
leads us to believe that a direct computation from their reflection
equations should be a starting point to obtain its classification.

After a systematic study of the functional equations we find that there are
two types of solutions for the $A_{n-1}^{(1)}$ models. We call of {\rm Type-I%
} the $K$-matrices with three free parameters and $n+2$ non-zero matrix
elements. These solutions were denoted by ${\Bbb K}_{ij}^{I}$ to emphasize
the non-zero element out of the diagonal and its symmetric, which results in 
$n(n-1)/2$ reflection $K$-matrices.

The {\rm Type-II} solutions are more interesting because their have many
free parameters. The $A_{n-1}^{(1)}$ models for $n$ odd, in addition to the 
{\rm Type-I} solutions, have $n$ {\rm Type-II} solutions with $2n-1$
non-zero matrix elements and $(2+[\frac{n}{2}])$ free parameters. It turns
out that for $n$ even we also have $n$ {\rm Type-II} solutions but half of
them are $K$-matrices with $2n$ non-zero matrix elements and $(2+\frac{n}{2}%
) $ free parameters, while the remaining ones have $2(n-1)$ non-zero matrix
elements with $(1+\frac{n}{2})$ free parameters.

The corresponding $K_{+}(u)$ are obtained from the isomorphism (\ref{re.8}).
Out of this classification we have the trivial solution $\left(
K_{-}=1,K_{+}=M\right) $ for these models. Thus we ended our discussion on
the reflection matrices for the vertex models associated with the $%
A_{n-1}^{(1)}$ affine Lie algebra.

To complete the classification for all non-exceptional Lie algebras we still
have to consider the vertex models associated with the $B_{n}^{(1)}$, $%
C_{n}^{(1)}$, $D_{n}^{(1)}$, $A_{2n}^{(2)}$ and $A_{2n-1}^{(2)}$ Lie
algebras.

{\bf Acknowledgment:} This work was supported in part by Funda\c{c}\~{a}o de
Amparo \`{a} Pesquisa do Estado de S\~{a}o Paulo--FAPESP--Brasil and by
Conselho Nacional de Desenvol\-{}vimento--CNPq--Brasil.

\appendix

\section{The $A_{1}^{(1)}$ Reflection K-Matrices}

This is a very special case among the $A_{n-1}^{(1)}$ models. We note that
there is only one general $K$-matrix \ with $4$ non-zero matrix elements %
\cite{deVega, GZ}. From the {\rm Type-IIa} $\ $solutions (\ref{second.10})
or from the {\rm Type-I} solutions (\ref{first.4}) it is the ${\Bbb K}_{12}$
matrix 
\[
{\Bbb K}_{12}^{I}=\left( 
\begin{array}{cc}
f_{11}(u) & h_{12}(u) \\ 
h_{21}(u) & {\rm e}^{2u}f_{11}(-u)%
\end{array}%
\right) 
\]%
Although there is no constraint equation in this case, the regular condition
(\ref{re.10}) has fixed in three the number of free parameters, in agreement
with all {\rm Type-I} reflection $K$-matrices.

\section{The A$_{2}^{(1)}$ Reflection K-Matrices}

This is also a special case because it has only the {\rm Type-I} solutions $%
{\Bbb K}_{12}^{I}$, ${\Bbb K}_{13}^{I}$ and ${\Bbb K}_{23}^{I}$. From (\ref%
{first.4}) we have 
\begin{eqnarray}
{\Bbb K}_{12}^{I} &=&f_{11}(u)E_{11}+{\rm e}%
^{2u}f_{11}(-u)E_{22}+h_{12}(u)E_{12}+h_{21}(u)E_{21}+{\cal X}_{3}(u)E_{33} 
\nonumber \\
&=&\left( 
\begin{array}{ccc}
f_{11}(u) & h_{12}(u) & 0 \\ 
h_{21}(u) & {\rm e}^{2u}f_{11}(-u) & 0 \\ 
0 & 0 & {\cal X}_{3}(u)%
\end{array}%
\right) ,  \label{b.1}
\end{eqnarray}%
with the four parameters $\beta _{11},\beta _{12},\beta _{21}$ and $\beta
_{33}$ satisfied the constraint equation%
\begin{equation}
\beta _{12}\beta _{21}=\left( \beta _{33}-\beta _{11}-2\right) \left( \beta
_{33}+\beta _{11}-2\right) .  \label{b.2}
\end{equation}%
Two diagonal solutions are derived from (\ref{b.1}) due to this constraint
equation%
\begin{eqnarray}
\lim_{\beta _{33}\rightarrow -\beta _{11}+2}{\cal X}_{3}(u) &=&{\rm e}%
^{2u}f_{11}(-u)  \nonumber \\
&\Rightarrow &D_{1}={\rm diag}(f_{11}(u),{\rm e}^{2u}f_{11}(-u),{\rm e}%
^{2u}f_{11}(-u))  \label{b.3}
\end{eqnarray}%
and%
\begin{eqnarray}
\lim_{\beta _{33}\rightarrow \beta _{11}+2}{\cal X}_{3}(u) &=&{\rm e}%
^{2u}f_{11}(u)  \nonumber \\
&\Rightarrow &D_{2}={\rm diag}(f_{11}(u),{\rm e}^{2u}f_{11}(-u),{\rm e}%
^{2u}f_{11}(u))  \label{b.4}
\end{eqnarray}%
The matrix ${\Bbb K}_{13}^{I}$ is also given by (\ref{first.4})%
\begin{eqnarray}
{\Bbb K}_{13}^{I} &=&f_{11}(u)E_{11}+{\rm e}%
^{2u}f_{11}(-u)E_{33}+h_{13}(u)E_{13}+h_{31}(u)E_{31}+{\cal Y}%
_{2}^{(1)}(u)E_{22}  \nonumber \\
&=&\left( 
\begin{array}{ccc}
f_{11}(u) & 0 & h_{13}(u) \\ 
0 & {\cal Y}_{2}^{(1)}(u) & 0 \\ 
h_{31}(u) & 0 & {\rm e}^{2u}f_{11}(-u)%
\end{array}%
\right) ,  \label{b.5}
\end{eqnarray}%
but now the constraint equation is%
\begin{equation}
\beta _{13}\beta _{31}=\left( \beta _{22}+\beta _{11}-2\right) \left( \beta
_{22}-\beta _{11}\right) ,  \label{b.6}
\end{equation}%
and the corresponding diagonal reductions are 
\begin{eqnarray}
\lim_{\beta _{22}\rightarrow -\beta _{11}+2}{\cal Y}_{2}^{(1)}(u) &=&{\rm e}%
^{2u}f_{11}(-u)  \nonumber \\
D_{3} &=&{\rm diag}(f_{11}(u),{\rm e}^{2u}f_{11}(-u),{\rm e}^{2u}f_{11}(-u))
\label{b.7}
\end{eqnarray}%
and%
\begin{eqnarray}
\lim_{\beta _{22}\rightarrow \beta _{11}}{\cal Y}_{2}^{(1)}(u) &=&f_{11}(u) 
\nonumber \\
D_{4} &=&{\rm diag}(f_{11}(u),f_{11}(u),{\rm e}^{2u}f_{11}(-u))  \label{b.8}
\end{eqnarray}%
For the solution named ${\Bbb K}_{23}^{I}$ we recall the equation (\ref%
{first.1}) with $i=2$ and $j=3$%
\begin{eqnarray}
{\Bbb K}_{23}^{I} &=&f_{22}(u)E_{22}+{\rm e}%
^{2u}f_{ii}(-u)E_{33}+h_{23}(u)E_{23}+h_{32}(u)E_{32}+{\cal Z}_{2}(u)E_{11} 
\nonumber \\
&=&\left( 
\begin{array}{ccc}
{\cal Z}_{2}(u) & 0 & 0 \\ 
0 & f_{22}(u) & h_{23}(u) \\ 
0 & h_{32}(u) & {\rm e}^{2u}f_{22}(-u)%
\end{array}%
\right) ,  \label{b.9}
\end{eqnarray}%
with the constraint%
\begin{equation}
\beta _{23}\beta _{32}=\left( \beta _{11}+\beta _{22}\right) \left( \beta
_{11}-\beta _{22}\right)  \label{b.10}
\end{equation}%
we get more two diagonal solutions%
\begin{eqnarray}
\lim_{\beta _{22}\rightarrow -\beta _{11}}{\cal Z}_{2}(u) &=&f_{22}(-u) 
\nonumber \\
D_{5} &=&{\rm diag}(f_{22}(-u),f_{22}(u),{\rm e}^{2u}f_{22}(-u))
\label{b.11}
\end{eqnarray}%
and%
\begin{eqnarray}
\lim_{\beta _{22}\rightarrow \beta _{11}}{\cal Z}_{2}(u) &=&f_{22}(u) 
\nonumber \\
D_{6} &=&{\rm diag}(f_{22}(u),f_{22}(u),{\rm e}^{2u}f_{22}(-u))  \label{b.12}
\end{eqnarray}

Due to the constraint equations these reflection $K$-matrices have only
three free parameters and the corresponding diagonal solutions have only one
free parameter.

Here we observe that only four of these diagonal solutions are independents
because $D_{6}=D_{4}$ and $D_{3}=D_{1}$. Here we also note that the
solutions $D_{1}$ and $D_{4}$ are the diagonal solutions derived by the
first time in \cite{deVega} and ${\Bbb K}_{13}^{I}$ is the non-diagonal
solution derived in \cite{Abad}.

In a certain sense these solution are particular because they do not reveal
us all properties shared by the regular $A_{n-1}^{(1)}$ reflection $K$%
-matrices for $n$ odd. Before we consider the next odd case, let us consider
the case $n=4.$

\section{The A$_{3}^{(1)}$ Reflection K-Matrices}

In this case the structure of the general solution begins to appear but it
is still particular because half of the {\rm Type-II} solutions are {\rm %
Type-I} solutions.

The ${\Bbb K}_{1j}$ matrices for the {\rm Type-I} solutions are given by (%
\ref{first.4}). For ${\Bbb K}_{12}^{I}$ we get 
\begin{equation}
{\Bbb K}_{12}^{I}=\left( 
\begin{array}{cccc}
f_{11}(u) & h_{12}(u) & 0 & 0 \\ 
h_{21}(u) & {\rm e}^{2u}f_{11}(-u) & 0 & 0 \\ 
0 & 0 & {\cal X}_{3}(u) & 0 \\ 
0 & 0 & 0 & {\cal X}_{3}(u)%
\end{array}%
\right)  \label{c.1}
\end{equation}%
with the constraint%
\begin{equation}
\beta _{12}\beta _{21}=\left( \beta _{33}+\beta _{11}-2\right) \left( \beta
_{33}-\beta _{11}-2\right)  \label{c.2}
\end{equation}%
For ${\Bbb K}_{13}^{I}$ we have%
\begin{equation}
{\Bbb K}_{13}^{I}=\left( 
\begin{array}{cccc}
f_{11}(u) & 0 & h_{13}(u) & 0 \\ 
0 & {\cal Y}_{2}^{(1)}(u) & 0 & 0 \\ 
h_{31}(u) & 0 & {\rm e}^{2u}f_{11}(-u) & 0 \\ 
0 & 0 & 0 & {\cal X}_{4}(u)%
\end{array}%
\right)  \label{c.4}
\end{equation}%
with the constraint%
\begin{equation}
\beta _{13}\beta _{31}=\left( \beta _{44}+\beta _{11}-2\right) \left( \beta
_{44}-\beta _{11}-2\right) =\left( \beta _{22}+\beta _{11}-2\right) \left(
\beta _{22}-\beta _{11}\right)  \label{c.5}
\end{equation}%
The ${\Bbb K}_{14}^{I}$ matrix is%
\begin{equation}
{\Bbb K}_{14}^{I}=\left( 
\begin{array}{cccc}
f_{11}(u) & 0 & 0 & h_{14}(u) \\ 
0 & {\cal Y}_{2}^{(1)}(u) & 0 & 0 \\ 
0 & 0 & {\cal Y}_{2}^{(1)}(u) & 0 \\ 
h_{41}(u) & 0 & 0 & {\rm e}^{2u}f_{11}(-u)%
\end{array}%
\right)  \label{c.7}
\end{equation}%
with

\bigskip

\begin{equation}
\beta _{14}\beta _{41}=\left( \beta _{22}+\beta _{11}-2\right) \left( \beta
_{22}-\beta _{11}\right)  \label{c.8}
\end{equation}%
The remaining {\rm Type-I} $K$-matrices are given by (\ref{first.1}). For $%
{\Bbb K}_{23}^{I}$ we get%
\begin{equation}
{\Bbb K}_{23}^{I}=\left( 
\begin{array}{cccc}
{\cal Z}_{2}(u) & 0 & 0 & 0 \\ 
0 & f_{22}(u) & h_{23}(u) & 0 \\ 
0 & h_{32}(u) & {\rm e}^{2u}f_{22}(-u) & 0 \\ 
0 & 0 & 0 & {\rm e}^{2u}{\cal Z}_{2}(u)%
\end{array}%
\right)  \label{c.10}
\end{equation}%
with constraint 
\begin{equation}
\beta _{23}\beta _{32}=\left( \beta _{11}+\beta _{22}\right) \left( \beta
_{11}-\beta _{22}\right)  \label{c.11}
\end{equation}%
For ${\Bbb K}_{24}^{I}$ we have%
\begin{equation}
{\Bbb K}_{24}^{I}=\left( 
\begin{array}{cccc}
{\cal Z}_{2}(u) & 0 & 0 & 0 \\ 
0 & f_{22}(u) & 0 & h_{24}(u) \\ 
0 & 0 & {\cal Y}_{3}^{(2)}(u) & 0 \\ 
0 & h_{42}(u) & 0 & {\rm e}^{2u}f_{22}(-u)%
\end{array}%
\right)  \label{c.13}
\end{equation}%
with the constraint 
\begin{equation}
\beta _{24}\beta _{42}=\left( \beta _{11}+\beta _{22}\right) \left( \beta
_{11}-\beta _{22}\right) =\left( \beta _{33}+\beta _{22}-2\right) \left(
\beta _{33}-\beta _{22}\right)  \label{c.14}
\end{equation}%
and finally for ${\Bbb K}_{34}^{I}$%
\begin{equation}
{\Bbb K}_{34}^{I}=\left( 
\begin{array}{cccc}
{\cal Z}_{3}(u) & 0 & 0 & 0 \\ 
0 & {\cal Z}_{3}(u) & 0 & 0 \\ 
0 & 0 & f_{33}(u) & h_{34}(u) \\ 
0 & 0 & h_{34}(u) & {\rm e}^{2u}f_{33}(-u)%
\end{array}%
\right)  \label{c.16}
\end{equation}%
with 
\begin{equation}
\beta _{34}\beta _{43}=\left( \beta _{11}+\beta _{33}\right) \left( \beta
_{11}-\beta _{33}\right)  \label{c.17}
\end{equation}

For the {\rm Type-IIa} solutions we get from (\ref{second.10}) more two $K$%
-matrices 
\begin{equation}
{\Bbb K}_{12}^{II}=\left( 
\begin{array}{cccc}
f_{11}(u) & h_{12}(u) & 0 & 0 \\ 
h_{21}(u) & {\rm e}^{2u}f_{11}(-u) & 0 & 0 \\ 
0 & 0 & {\rm e}^{2u}f_{11}(-u) & {\rm e}^{u}h_{34}(u) \\ 
0 & 0 & {\rm e}^{u}h_{43}(u) & {\rm e}^{2u}f_{11}(u)%
\end{array}%
\right)  \label{c.19}
\end{equation}%
with the eight non-zero elements satisfying a constraint equation 
\begin{equation}
\beta _{12}\beta _{21}=\beta _{34}\beta _{43}.  \label{c.20}
\end{equation}%
The another $K$-matrix is given by%
\begin{equation}
{\Bbb K}_{14}^{II}=\left( 
\begin{array}{cccc}
f_{11}(u) & 0 & 0 & h_{14}(u) \\ 
0 & f_{11}(u) & h_{23}(u) & 0 \\ 
0 & h_{32}(u) & {\rm e}^{2u}f_{11}(-u) & 0 \\ 
h_{41}(u) & 0 & 0 & {\rm e}^{2u}f_{11}(-u)%
\end{array}%
\right)  \label{c.21}
\end{equation}%
with the a constraint equation%
\begin{equation}
\beta _{14}\beta _{41}=\beta _{23}\beta _{32}  \label{c.22}
\end{equation}%
Note that both {\rm Type-II} solutions (\ref{c.19}) and (\ref{c.21}) have
four free parameters.

Next, we can solve these constraint equations to derive eighteen diagonal
solutions. Using the following reductions for the scalar functions ${\cal X}%
_{j+1}(u)$, ${\cal Y}_{l}^{(i)}(u)$ and ${\cal Z}_{i}(u)$%
\begin{eqnarray}
\lim_{\beta _{j+1\ j+1}\rightarrow -\beta _{11}+2}{\cal X}_{j+1}(u) &=&{\rm e%
}^{2u}f_{11}(-u),\quad \lim_{\beta _{j+1\ j+1}\rightarrow \beta _{11}+2}%
{\cal X}_{j+1}(u)={\rm e}^{2u}f_{11}(u),  \nonumber \\
\quad \lim_{\beta _{ll}\rightarrow -\beta _{ii}+2}{\cal Y}_{l}^{(i)}(u) &=&%
{\rm e}^{2u}f_{ii}(-u),\quad \quad \lim_{\beta _{ll}\rightarrow \beta _{ii}}%
{\cal Y}_{l}^{(i)}(u)=f_{ii}(u),  \nonumber \\
\lim_{\beta _{11}\rightarrow -\beta _{ii}}{\cal Z}_{i}(u)
&=&f_{ii}(-u),\quad \lim_{\beta _{11}\rightarrow \beta _{ii}}{\cal Z}%
_{i}(u)=f_{ii}(u).
\end{eqnarray}%
we can see that only half of these diagonal solutions are independents:%
\begin{eqnarray}
D_{1} &=&{\rm diag}(f(u),{\rm e}^{2u}f(-u),{\rm e}^{2u}f(-u),{\rm e}%
^{2u}f(-u)),  \nonumber \\
D_{2} &=&{\rm diag}(f(u),{\rm e}^{2u}f(-u),{\rm e}^{2u}f(u),{\rm e}%
^{2u}f(u)),  \nonumber \\
D_{3} &=&{\rm diag}(f(u),f(u),{\rm e}^{2u}f(-u),{\rm e}^{2u}f(-u)), 
\nonumber \\
D_{4} &=&{\rm diag}(f(u),{\rm e}^{2u}f(-u),{\rm e}^{2u}f(-u),{\rm e}%
^{2u}f(u)),  \nonumber \\
D_{5} &=&{\rm diag}(f(u),f(u),{\rm e}^{2u}f(-u),{\rm e}^{2u}f(u)),  \nonumber
\\
D_{6} &=&{\rm diag}(f(u),f(u),f(u),{\rm e}^{2u}f(-u)),  \nonumber \\
D_{7} &=&{\rm diag}(f(-u),f(u),{\rm e}^{2u}f(-u),{\rm e}^{2u}f(-u)), 
\nonumber \\
D_{8} &=&{\rm diag}(f(-u),f(u),f(u),{\rm e}^{2u}f(-u)),  \nonumber \\
D_{9} &=&{\rm diag}(f(-u),f(-u),f(u),{\rm e}^{2u}f(-u)),
\end{eqnarray}%
where we have used a compact notation for the functions $f_{ii}(u)$ 
\begin{equation}
f_{ii}(u)\equiv f(u)=\beta (e^{u}-1)+1
\end{equation}%
where $\beta $ is the free parameter.

\section{\protect\bigskip The A$_{4}^{(1)}$ Type-II Reflection K-Matrices}

Here we will only write explicitly the five {\rm Type-II} \ solutions and
their constraint equations for the $A_{4}^{(1)}$ model. They have nine
non-zero matrix elements and four free parameters:

\begin{eqnarray}
{\Bbb K}_{12}^{II} &=&\left( 
\begin{array}{ccccc}
f_{11}(u) & h_{12}(u) & 0 & 0 & 0 \\ 
h_{21}(u) & {\rm e}^{2u}f_{11}(-u) & 0 & 0 & 0 \\ 
0 & 0 & {\rm e}^{2u}f_{11}(-u) & 0 & {\rm e}^{u}h_{35}(u) \\ 
0 & 0 & 0 & {\cal X}_{4}(u) & 0 \\ 
0 & 0 & {\rm e}^{u}h_{53}(u) & 0 & {\rm e}^{2u}f_{11}(u)%
\end{array}%
\right) ,  \nonumber \\
\beta _{12}\beta _{21} &=&\beta _{35}\beta _{53}=(\beta _{44}+\beta
_{11}-2)(\beta _{44}-\beta _{11}-2),  \label{d.1}
\end{eqnarray}%
\begin{eqnarray}
{\Bbb K}_{14}^{II} &=&\left( 
\begin{array}{ccccc}
f_{11}(u) & 0 & 0 & h_{14}(u) & 0 \\ 
0 & f_{11}(u) & h_{23}(u) & 0 & 0 \\ 
0 & h_{32}(u) & {\rm e}^{2u}f_{11}(-u) & 0 & 0 \\ 
h_{41}(u) & 0 & 0 & {\rm e}^{2u}f_{11}(-u) & 0 \\ 
0 & 0 & 0 & 0 & {\cal X}_{5}(u)%
\end{array}%
\right) ,  \nonumber \\
\beta _{14}\beta _{41} &=&\beta _{23}\beta _{32}=(\beta _{55}+\beta
_{11}-2)(\beta _{55}-\beta _{11}-2),  \label{d.2}
\end{eqnarray}%
\begin{eqnarray}
{\Bbb K}_{13}^{II} &=&\left( 
\begin{array}{ccccc}
f_{11}(u) & 0 & h_{13}(u) & 0 & 0 \\ 
0 & {\cal X}_{2}(u) & 0 & 0 & 0 \\ 
h_{31}(u) & 0 & {\rm e}^{2u}f_{11}(-u) & 0 & 0 \\ 
0 & 0 & 0 & {\rm e}^{2u}f_{11}(-u) & {\rm e}^{u}h_{45}(u) \\ 
0 & 0 & 0 & {\rm e}^{u}h_{54}(u) & {\rm e}^{2u}f_{11}(u)%
\end{array}%
\right) ,  \nonumber \\
\beta _{13}\beta _{31} &=&\beta _{45}\beta _{54}=(\beta _{22}+\beta
_{11}-2)(\beta _{22}-\beta _{11}),  \label{d.3}
\end{eqnarray}%
\begin{eqnarray}
{\Bbb K}_{15}^{II} &=&\left( 
\begin{array}{ccccc}
f_{11}(u) & 0 & 0 & 0 & h_{15}(u) \\ 
0 & f_{11}(u) & 0 & h_{24}(u) & 0 \\ 
0 & 0 & {\cal X}_{3}(u) & 0 & 0 \\ 
0 & h_{42}(u) & 0 & {\rm e}^{2u}f_{11}(-u) & 0 \\ 
h_{51}(u) & 0 & 0 & 0 & {\rm e}^{2u}f_{11}(-u)%
\end{array}%
\right) ,  \nonumber \\
\beta _{15}\beta _{51} &=&\beta _{24}\beta _{42}=(\beta _{33}+\beta
_{11}-2)(\beta _{33}-\beta _{11}),  \label{d.4}
\end{eqnarray}%
\begin{eqnarray}
{\Bbb K}_{25}^{II} &=&\left( 
\begin{array}{ccccc}
{\cal Z}_{2}(u) & 0 & 0 & 0 & 0 \\ 
0 & f_{11}(u) & 0 & 0 & h_{25}(u) \\ 
0 & 0 & f_{11}(u) & h_{34}(u) & 0 \\ 
0 & 0 & h_{43}(u) & {\rm e}^{2u}f_{11}(-u) & 0 \\ 
0 & h_{52}(u) & 0 & 0 & {\rm e}^{2u}f_{11}(-u)%
\end{array}%
\right) ,  \nonumber \\
\beta _{25}\beta _{52} &=&\beta _{34}\beta _{43}=(\beta _{11}+\beta
_{22})(\beta _{11}-\beta _{22}).  \label{d.5}
\end{eqnarray}

The corresponding diagonal solutions are also one-parameter solutions.

\end{document}